\documentclass[10pt]{iopart}

\usepackage{tabularx}
\expandafter\let\csname equation*\endcsname\relax
\expandafter\let\csname endequation*\endcsname\relax
\usepackage{amsmath}
\usepackage{graphicx}
\usepackage[final]{hyperref} 
\hypersetup{
	colorlinks=true,       
	linkcolor=blue,        
	citecolor=blue,        
	filecolor=magenta,     
	urlcolor=blue         
}
\usepackage{fancyhdr}
\fancypagestyle{specialfooter}{%
	\fancyhf{}
	
	\fancyfoot[C]{Distribution A: Approved for public release; Distribution unlimited}
}

\begin{document}

\title[A comparison of Fourier and POD mode decomposition methods for high-speed Hall thruster video]{A comparison of Fourier and POD mode decomposition methods for high-speed Hall thruster video}

\author{J. W. Brooks \& M. S. McDonald}

\address{Naval Research Laboratory, Washington DC}
\ead{john.brooks@nrl.navy.mil \& michael.mcdonald@nrl.navy.mil}
\vspace{10pt}

\author{A. A. Kaptanoglu}

\address{University of Washington, Seattle WA}
\ead{akaptano@uw.edu}
\vspace{10pt}

\begin{abstract}
Hall thrusters are susceptible to large-amplitude plasma oscillations that impact thruster performance and lifetime and are also difficult to model.  High-speed cameras are a popular tool to study these dynamics due to their spatial resolution and are a popular, nonintrusive complement to in-situ probes.  High-speed video of thruster oscillations can be isolated (decomposed) into coherent structures (modes) with algorithms that help us better understand the evolution and interactions of each.  This work provides an introduction, comparison, and step-by-step tutorial on established Fourier and newer Proper Orthogonal Decomposition (POD) algorithms as applied to high-speed video of the unshielded H6 6-kW laboratory model Hall thruster.  From this dataset, both sets of algorithms identify and characterize $m=0$ and $m>0$ modes in the discharge channel and cathode regions of the thruster plume, as well as mode hopping between the $m=3$ and $m=4$ rotating spokes in the channel. The Fourier methods are ideal for characterizing linear modal structures and also provide intuitive dispersion relationships.  By contrast, the POD method tailors a basis set using energy minimization techniques that better captures the nonlinear nature of these structures and with a simpler implementation.  Together, the Fourier and POD methods provide a more complete toolkit for studying Hall thruster plasma instabilities and mode dynamics.  Specifically, we recommend first applying POD first to quickly identify the nature and location of global dynamics and then using Fourier methods to isolate dispersion plots and other wave-based physics.  
\end{abstract}

%
\maketitle
%
\ioptwocol
%

\thispagestyle{specialfooter} 

\section{\label{sec:intro}Introduction}

The Hall thruster (HT) is a standard spacecraft electric propulsion system that uses crossed electric and magnetic fields $\left ( \mathbf{E} \times \mathbf{B} \right)$ to ionize and accelerate propellant~\cite{goebel2008,boeuf2017}.  Their high specific impulse and technological maturity make them ideal for long duration satellite station keeping and time-insensitive missions, such as orbit raising, with thousands currently in orbit and more planned.  However, HTs experience  anomalous electron transport across their magnetic field that is not sufficiently understood to permit fully predictive thruster models, motivating continued improvement of diagnostics for model validation.  This lack of validation is especially important for increased qualification by simulation for new thruster designs at ever-higher power, such as the N30 or X3 nested Hall thrusters, where full life qualification in a ground test facility would be prohibitively expensive for 10s-100s kW thrusters~\cite{florenz2012,hall2017}. 

Anomalous electron transport has been strongly linked to plasma oscillations in HTs~\cite{choueiri2001} in both the thruster discharge channel~\cite{parker2010, mcdonald2011segAnode} and cathode regions~\cite{jorns2020}.  Long-wavelength azimuthal oscillations, the focus of this work, are typically characterized in a Fourier representation with integer mode numbers, $m$, including the azimuthally uniform $m=0$ “breathing” modes in HT channels~\cite{sekerak2016,dale2019} and cathodes~\cite{goebel2007,georgin2019}, $m>0$ azimuthally rotating “spoke” modes in the channel, and a $m=1$ counter rotating “anti-drift” modes around the cathode\cite{jorns2014}.  The rotating spoke channel modes, first observed in an early Hall accelerator by Janes and Lowder~\cite{janes1966}, are now known to be ubiquitous in unshielded thrusters\cite{mcdIEEE-TPSrotspoke} though they may be less common in magnetically shielded thrusters~\cite{jorns2014, bairdthesis}.  Cathode $m=1$ anti-drift modes have been previously observed in both shielded and unshielded versions of the H6 Hall thruster~\cite{jorns2014} (the unshielded H6 is studied in this work).  All of these modes are historically characterized by one or more in-situ diagnostics, but the localized nature of these diagnostics both perturb the plasma and make it difficult to relate individual measurements to global mode structures.  

High-speed imaging (HSI) has become a popular complementary diagnostic to \textit{in situ} probes and has proven itself well suited to characterizing mode dynamics~\cite{parker2010,mcdonald2013,jorns2014}.  Its popularity is due to its ease of use, nonintrusive nature, high speed (100s of kHz), and fine spatial resolution of order millimeters per pixel.  HSI was first used on HTs to relate image brightness to plasma density oscillations measured by electrical probes~\cite{darnon1997}.  This and later work has led to the use of pixel light intensity as a proxy for discharge current~\cite{hara2014}.  

To isolate and characterize individual mode dynamics from HSI, various post-processing algorithms have been developed.  Early work subtracted the time-average from each pixel to reveal multiple $m>1$ modes in the H6 Hall thruster channel~\cite{mcdIEEE-TPSrotspoke}.  Later, Fourier techniques were used with azimuthal binning to isolate the frequency spectrum associated with individual channel modes~\cite{mcdonald2011spokeUbiquity}.   Fourier methods have also been used to provide azimuthal dispersion plots within the CHT (cylindrical Hall thruster)~\cite{parker2010} and H6~\cite{mcdonald2013} thruster channels.    A phase-based Fourier analysis was developed to isolate the spatial structure of the $m=1$ cathode mode on the H6 thruster~\cite{jorns2014} and $m>0$  channel and cathode modes on the HERMeS thruster~\cite{bairdthesis}.  An alternative Fourier-based visualization method, Cross-Spectral-Density (CSD), was developed on the CHT thruster~\cite{romadanov2019}.  

The Fourier methods remain the most established method for isolating and analyzing Hall thruster plasma oscillations, but they have several disadvantages.  First, their linear sine/cosine bases are not ideal for nonlinear features.  Second, Fourier methods require several preprocessing steps for high-speed video; this added complexity requires additional computational overhead and makes it more difficult to study multi-dimensional dynamics.  

An alternative to Fourier methods are Singular Value Decomposition (SVD) based algorithms~\cite{brunton2019}. The chief advantage of SVD is that is tailors custom bases for each dataset based on the energy contribution of the coherent dynamics instead of using a presumed basis (e.g. Fourier's sines and cosines).  This facilitates improved characterization and reconstruction of nonlinear dynamics without making any physical assumptions of the system. SVD also requires minimal preprocessing compared with Fourier methods.  

The most notable and likely simplest SVD algorithm is Proper Orthogonal Decomposition (POD) which is extensively used in fluid mechanics~\cite{Benner2015siamreview,Taira2017aiaa,brunton2019}.  A classic POD application is to decompose fluid vortex shedding around a body (e.g. a cylinder) into discrete modes~\cite{noack2003hierarchy,oudheusden2005}.  POD has also been used in plasma physics to characterize plasma oscillations~\cite{dudok1994,levesque2013,vanMilligen14,hansen2015numerical,kaptanoglu2020physics} 
but is often referred to as Biorthogonal Decomposition (BD).  Very recently, POD has been used to characterize axial and azimuthal modes in Hall thruster high-speed video~\cite{desangles2020} and azimuthal modes in a hollow cathode plume~\cite{Becatti2021}.  More sophisticated SVD algorithms exist, most notably DMD (Dynamic Mode Decomposition)~\cite{schmid2010dynamic,Tu2014jcd} and have applications in active control, linear dynamics, and plasma physics~\cite{taylor2018dynamic,Sasaki2019,kaptanoglu2020}.

The goal of this work is to compare Fourier and POD techniques as applied to Hall thruster high-speed imaging in a tutorial format.  To this end, this work analyzes a high-speed video recording of the unshielded H6 Hall thruster plume and provides a step-by-step explanation of algorithm's implementation and results.  This work starts by introducing the H6 thruster, high-speed video dataset, and video preprocessing in Section~\ref{sec:setup}.  Section~\ref{sec:Fourier} discusses several established Fourier mode analysis methods and, when applied to the H6 dataset, identifies simultaneous cathode and channel modes in addition to mode hopping within the channel.   Section~\ref{sec:SVDmethod} introduces the SVD algorithm and its most common implementation: Proper Orthogonal Decomposition (POD).
When applied to the H6 dataset, POD identifies the same mode behavior and provide several improvements over the Fourier methods.   When used together, POD and Fourier methods provide a more complete toolkit for identifying and the isolating global mode dynamics.  The code and dataset used for this work are available online~\cite{zenodo_2021}.  

\section{\label{sec:setup}Experimental setup}

This section covers the experimental hardware (thruster, facility, high-speed camera), the HSI dataset, and common HSI prepossessing techniques.  

\subsection{Hardware and dataset}

\begin{figure}
	\includegraphics[]{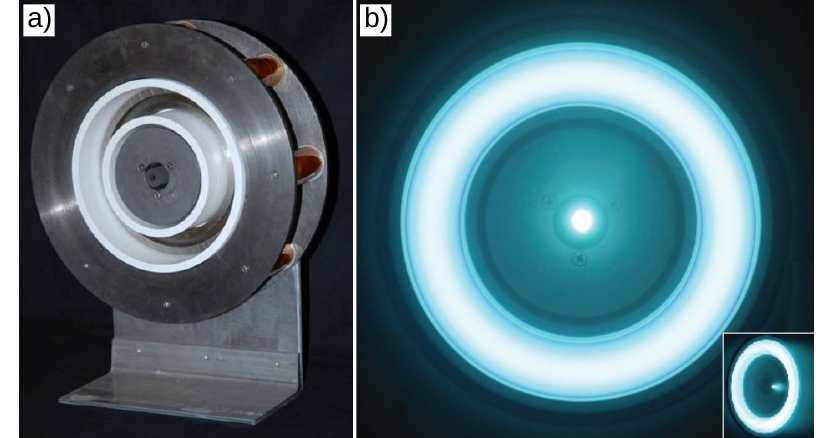}
	\caption{\label{fig:h6}  a) The unshielded H6 Hall thruster.  b) End-on-view of the thruster and the plasma in its channel (annular ring) and the cathode (center dot).  Figure reproduced here with permission from the author~\cite{brown2009}.}
\end{figure}

This work focuses on a single illustrative operating condition for the unshielded H6 Hall thruster.  The H6 is a laboratory model 6-kW thruster (see Figure~\ref{fig:h6}) with a design operating point of 300 V and 20 mg/s (where 1 mg/s $\approx$ 1 A) on xenon.  In this work, we focus on a 600 V and 10 mg/s condition that exhibits several simultaneous modes, most prominently an $m=1$ cathode spoke, a shared $m=0$ mode between the channel and cathode regions, and mode-hopping between $m=3$ and $m=4$ in the channel.   

Recordings took place in the University of Michigan Plasmadynamics and Electric Propulsion Laboratory's Large Vacuum Test Facility (LVTF) circa 2011.  
The LVTF is a cylindrical chamber 9 m long and 6 m in diameter and at the time was maintained at high vacuum by seven TM-1200 cryopumps with a combined pumping speed of 210,000 L/s on xenon.

A Photron SA5 FASTCAM high-speed camera placed 6.5~m axially downstream of the thruster exhaust imaged the plume through a quartz window with 152~x~192 resolution in monochrome (B\&W) at a framerate of 175 kHz.  The fastest mode observed was a 80 kHz cathode spoke, just below the 87.5 kHz camera Nyquist frequency.  The camera used a Nikon ED AF Nikkor 80-200 mm lens at its maximum aperture of f/2.8.   The bright, central cathode saturates several pixels, and the algorithms largely ignore these pixels.  

In preparation for video processing, any high-speed video dataset should be thought of as a 3D matrix of pixel measurements, $p(t,x,y)$, with dimensions of time, $t$, and Cartesian space, $x$ and $y$, and with lengths $N_t$, $N_x$, and $N_y$, respectively~\cite{mcdonald2013}. 

\subsection{\label{sec:preprocessing}Video preprocessing}

Before applying mode analysis algorithms, several preprocessing steps should first be considered.  This section outlines two prominent steps: 1) spatial centering and normalization and 2) amplitude normalization. 
Other preprocessing steps, not covered here, include masking and filtering which are useful in isolating specific spatial regions or dynamics, respectively. Please note that all of these steps are optional for POD.

\subsubsection{\label{sec:precond_spatial} Spatial identification and scaling}

This step identifies and the scales the spatial geometries in preparation for converting to polar coordinates, a requirement for Fourier analysis. This step is not required for POD.  To identify the thruster channel origin and radius, we fit an annular Gaussian function  

\begin{eqnarray}
	\eqalign{G&(x,y;x_0,y_0,r_0,a,w, G_0) \label{eq:annular_gaussian}\\
	&= a \, exp \left ( - \frac{1}{2} \left( \frac{ r (x,y;x_0,y_0) - r_0}{w}\right)^2  \right) + G_0 } 
\end{eqnarray}

to the time-average, $\bar{p}(x,y)$, of the high-speed video.  In this equation, $r_0$ is the radius of the channel, the radius at each pixel is

\begin{equation}
	r(x,y;x_0,y_0 )=\sqrt{ (x-x_0 )^2+(y-y_0 )^2 },
	\label{eq:radius}
\end{equation}
$x_0$ and $y_0$ are the center (origin) of the channel, $a$ is the amplitude, $w$ is related to the channel width, and $G_0$ is an offset.  

With these fit parameters solved, we next center and normalize the spatial coordinates to the channel radius (i.e. $x_{norm}=(x-x_0)/r_0$ and $y_{norm}=(y-y_0)/r_0$ ).  These normalized coordinates can then be optionally multiplied by the dimensioned channel radius to provide actual units to $x$ and $y$.  For this work, we remain with normalized coordinates.  

\begin{figure}
	\includegraphics{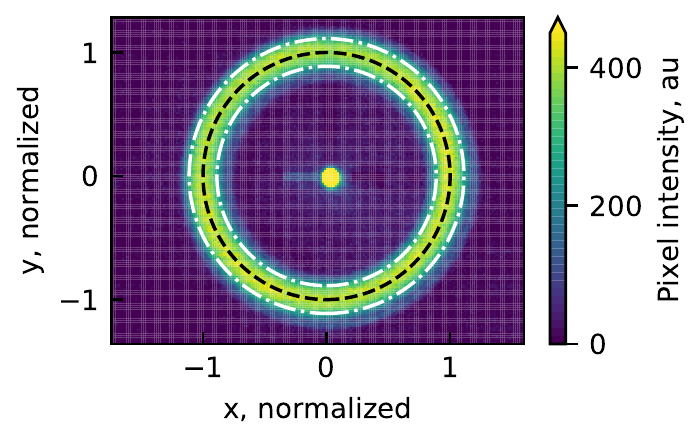}
	\caption{\label{fig:precond_spatial} An annular Gaussian function is fit to the time-averaged video and identifies the channel origin and radius.  The video's coordinates are then centered at the origin and normalized by the channel radius.   }
\end{figure}

Figure~\ref{fig:precond_spatial} shows the results of this step as applied to the time-averaged video, $\bar{p}(x,y)$, of our H6 dataset.  In this figure, the two regions with the most plasma dynamics, the annular channel and center cathode, are clearly visible.  The pixel intensity has raw integer units based on the camera's 12 bit depth.  The pixels associated with the cathode's center were allowed to saturate to better capture dynamics in the channel, and the pixels adjacent to the saturated pixels still capture the cathode's dynamics.  The channel radius, $r_0$, as identified by the Gaussian fit, is indicated with a black dashed line.  The channel edges, indicated with white dashed lines, are identified as approximately at $(r_0\pm w)/r_0$.  The $x$ and $y$ coordinates have been centered and normalized to the channel radius as described above.  

Alternatives to the annular Gaussian fit have been used in previous Hall thruster work.  McDonald~\cite{mcdonald2013} discussed both the Kasa~\cite{kasa1976} and Taubin~\cite{taubin1991} methods and recommended Taubin for cases where the entire channel annulus is not visible.   Another option previously used~\cite{romadanov2019} is a circle detection algorithm called a Hough transform.  While no one method is obviously superior over the others, we recommend using 1) the Gaussian fit presented here because the solved parameters are directly relatable to physical dimensions or 2) the Hough transform as it is available prewritten in many programming languages.  

\subsubsection{\label{sec:precond_amplitude} Video amplitude scaling}

This step scales the video's arbitrary intensity measurements to a more meaningful range.  
Common practice is to assume that oscillations in a video's amplitude is roughly linear with the plasma density oscillations~\cite{darnon1997} and that camera measurements are linear with light emission~\cite{vora1997}.  As we cannot scale the data to a definitive physical value, this section instead discusses several data normalization methods to provide better physical intuition of the oscillations.  This step is optional for both Fourier and POD methods.

Before normalizing, the first step is to subtract the time-averaged image (also known as AC coupling) 
from each frame of the raw video dataset to better isolate the oscillations.  

Next, we normalize the video amplitude in one of two ways: pixel-wise normalization or channel-average normalization.  The first method divides each pixel by its standard deviation in time, and this has the advantage of making the mode dynamics easier to visualize after mode decomposition.  Unfortunately, this method  artificially amplifies the oscillations at different spatial locations and makes quantifying global mode amplitudes untenable.  As an alternative, the second option divides each pixel by the average brightness within the channel, which allows the modes to be scaled as a percentage of the average channel brightness.  
While both methods are used in this work, channel-average normalization is default.  

\begin{figure}
	\includegraphics{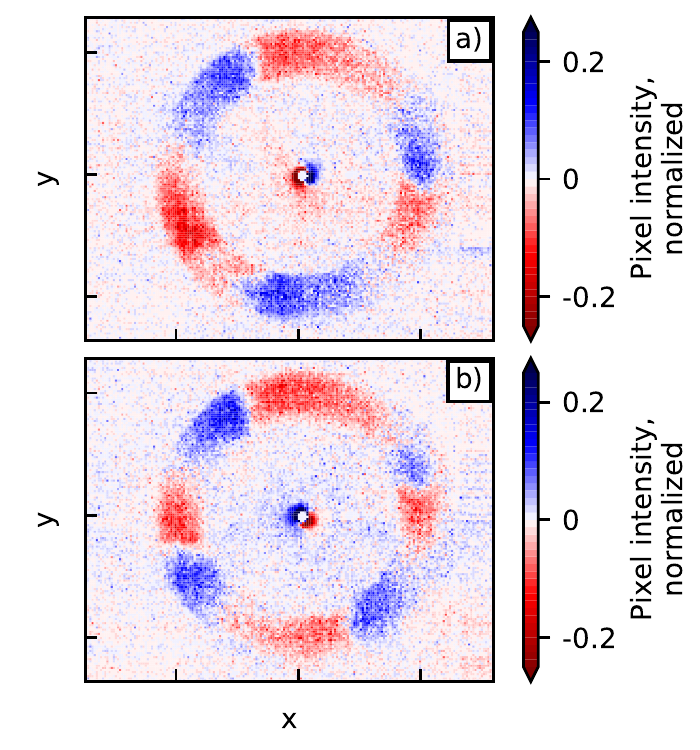}
	\caption{\label{fig:precond_amptliude} Two time snapshots, that have been AC coupled and normalized by the average channel brightness, reveal prominent mode structures.  a) At $t=22.9$~ms, an $m=3$ mode is dominant in the channel.  b) At $t=24.8$~ms, an $m=4$ mode is dominant in the channel.  Both snapshots show an $m=1$ mode around the cathode.   }
\end{figure}

Figure~\ref{fig:precond_amptliude} shows the results of AC coupling and channel normalization at two separate instances in time.  At $t=22.9$ ms (Figure~\ref{fig:precond_amptliude}a), the normalized video snapshot reveals a dominant 3-lobed azimuthal wave ($m=3$ mode) in the channel.  At $t=24.8$ ms (Figure~\ref{fig:precond_amptliude}b), a 4-lobed azimuthal wave ($m=4$ mode) is dominant.  The amplitudes of both modes are around 10\% of the average channel brightness.  Both snapshots also show an $m=1$ azimuthal wave around the cathode.

\section{\label{sec:Fourier}Fourier-based methods}

Fourier based algorithms are the most established methods for mode decomposition and identification in HT high-speed video~\cite{mcdonald2011,mcdonald2013,jorns2014,romadanov2019}.   This is because Fourier series' bases are periodic sines and cosines and are therefore ideal for characterizing wave-like oscillations, such as plasma waves.  In this section, we use Fourier analysis to characterize simultaneous $m=0$ and $m>0$ modes associated with the thruster discharge channel and cathode in addition to mode hopping between the $m=3$ and $m=4$ modes in the channel.

\subsection{Detecting modes}

\begin{figure*}
	\includegraphics{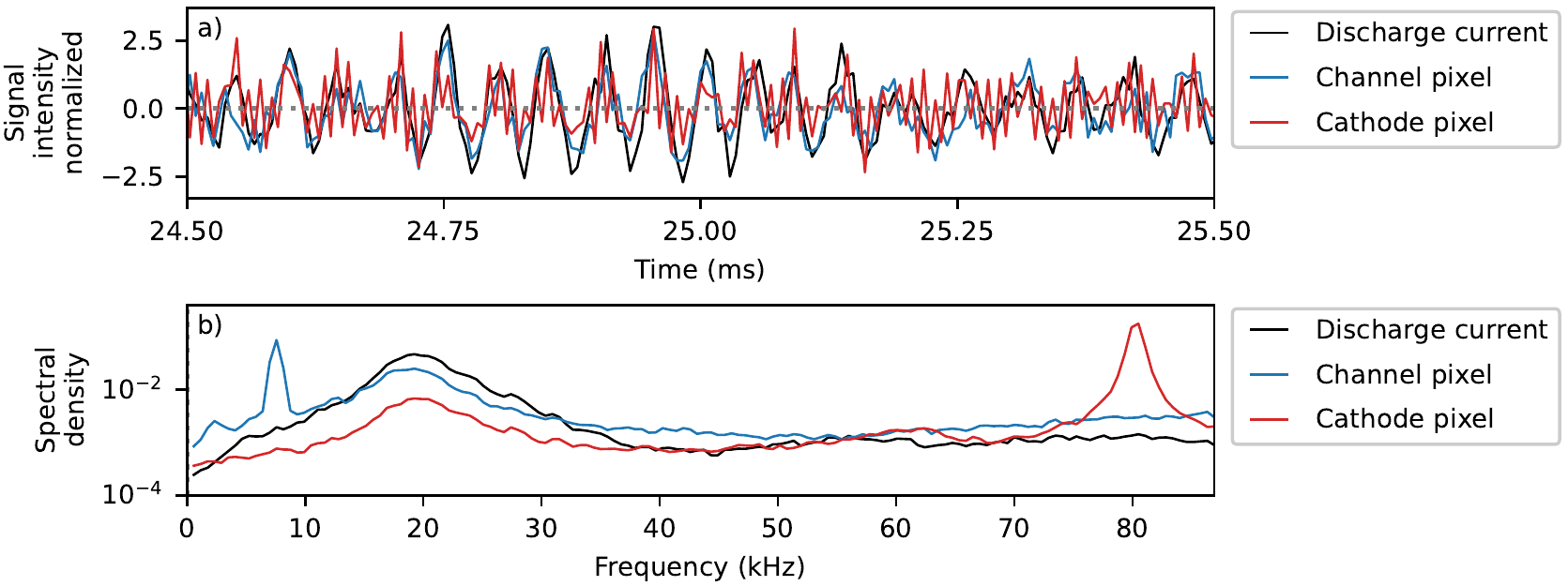}
	\caption{\label{fig:raw_data} Three signals are analyzed (a pixel in the thruster channel ($r/r_0=1$), a pixel adjacent to the thruster cathode ($r/r_0=0.1$), and the discharge current), and their power spectrum reveals the existence of several prominent modes (peaks).  a) The three time-series signals over 1~ms.  Each has been subtracted by their mean and divided by their standard deviation.  b)  Their power spectrums, calculated over 100~ms.    
	}
\end{figure*}

Mode dynamics in Hall thruster plasmas are typically characterized by oscillating waves with slowly evolving amplitudes and frequencies.  The easiest way to detect these modes is to apply FFT and Welch-averaged FFT~\cite{welch1967} algorithms to diagnostic measurements of signal pixels of the high-speed video.  Figure~\ref{fig:raw_data} shows an example of this applied to three signals from our dataset:  the discharge current flowing from the anode to cathode, a high-speed pixel centered in the channel, and a high-speed pixel adjacent to the cathode.  

Figure~\ref{fig:raw_data}a shows a 1~ms time window of the three signals and that they have similar oscillatory behavior.  Each signal has been AC coupled and normalized by its standard deviation for ease of comparison.  

Figure~\ref{fig:raw_data}b shows the Welch-averaged FFT of each signal over 100 ms, and the resulting power spectrum reveals prominent frequency peaks with each.  Most notable is a broad spectrum peak at 20 kHz that is common to all three signals.   Additional peaks at 7.5 kHz and 80 kHz are unique to the channel and cathode measurements, respectively; this uniqueness suggests that their dynamics are isolated to their respective regions.  The broader width of the 20 kHz peak suggests that its frequency is more erratic than the two narrower peaks.  The remainder of this paper will identify the modes associated with these three peaks and few less pronounced peaks.

\subsection{Azimuthal and radial binning}

Waves around the channel and cathode of a Hall thruster are primarily azimuthal~\cite{choueiri2001} with the approximate Fourier form

\begin{equation}
	p(t,\theta) \sim e^{i(m\theta - \omega t)}
	\label{eq:fourier}
\end{equation}
\noindent where $m$ is the integer azimuthal mode number, $\theta$ is the azimuthal angle, and $\omega = 2 \pi f$ is the angular frequency.  After scaling the video's coordinates and amplitudes (Section~\ref{sec:preprocessing}), the next step is convert the 3D video, $p(t,x,y)$, in Cartesian coordinates to a 2D video, $p(t,\theta)$, in polar coordinates so that it matches Eq.~\ref{eq:fourier}.  To do this, we first isolate the channel region (i.e. $0.9 < r/r_0 < 1.1$) with radial masking and discard the rest.  The radial dependence within this narrow region is assumed constant, and the radial coordinate is therefore dropped.

Next, we convert the unmasked $x$ and $y$ coordinates within the channel to the azimuthal coordinate with a four-quadrant inverse tangent function, $\theta(x,y)=\text{arctan2}(y,x)$.  The resulting data is azimuthally binned and averaged to provide a result with uniform azimuthal spacing.   

\begin{figure}
	\includegraphics{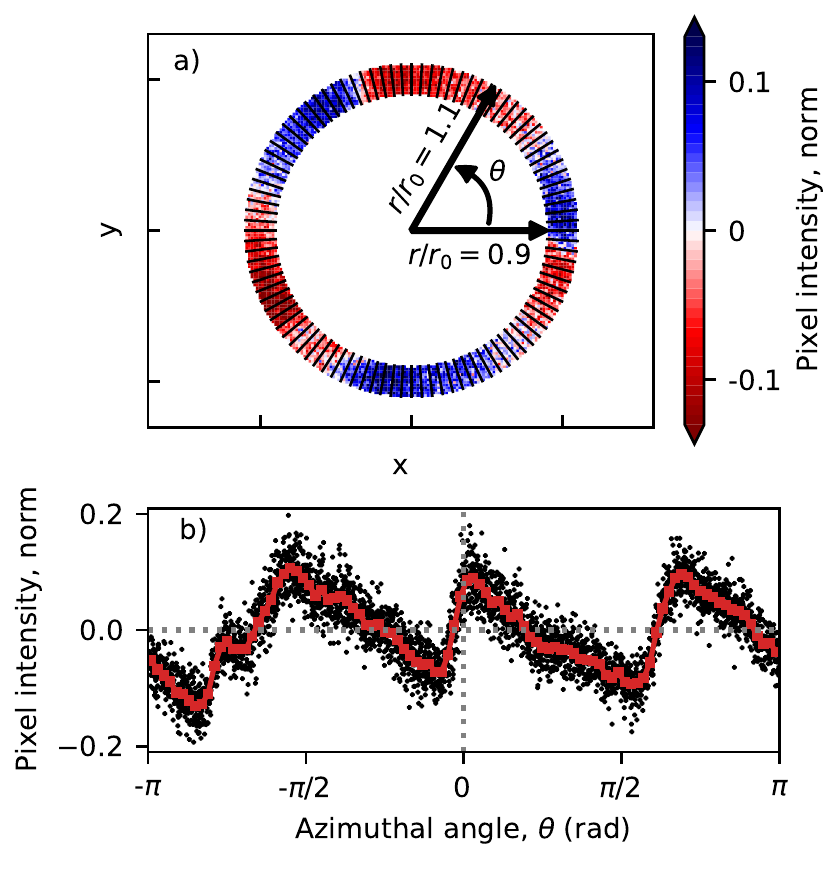}
	\caption{\label{fig:azim_binning} The azimuthal binning process for a single instant in time is shown.
		a) A radial mask isolates a narrow radial region within the channel, and the $N_\theta$=100 azimuthal bin boundaries are overlayed.  b) The average of each azimuthal bin (red) overlays the raw (black) data.   }
\end{figure}

Figure~\ref{fig:azim_binning} illustrates the azimuthal and radial binning for a single time snapshot ($t=22.9$ ms).  Figure~\ref{fig:azim_binning}a shows the radial mask isolating a narrow region within the channel and also the edges of $N_\theta=100$ azimuthal bins.  Figure~\ref{fig:azim_binning}b shows the unbinned values (black) overlayed by the bin-averaged result (red).  This shows a clear $m=3$  structure and also a few non-ideal features: a non-uniform spacing between peaks, a steepened waveform, and a weak fourth peak at 3$\pi$/4.   This process is repeated for each time step within $p(t,x,y)$, and the result is the 2D dataset, $p(t,\theta)$, within the channel. 

In this example, we applied the binning process to the channel region.  However, this same procedure could be applied to other radial regions, including the region around the cathode~\cite{jorns2014} or at radial slices between the cathode and channel.  Optionally, multiple radial slices could be made to provide a 3D dataset, $p(t, \theta, r)$.

\subsection{Azimuthal mode identification \label{sec:fft_mode_id}}

With $p(t,\theta)$ solved and matching the assumed Fourier form (Eq.~\ref{eq:fourier}), we can next identify the azimuthal modes within the channel.  To do this, we apply a 2D FFT, $\mathcal{F}_{t,\theta}$, in both time and $\theta$ to identify the mode numbers and their characteristic frequencies.  
This provides the 2D complex matrix,

\begin{eqnarray}
	P(\omega,m)=\mathcal{F}_{t,\theta} \bigl \{p(t,\theta) \bigr \},
	\label{eq:FFT_2D}
\end{eqnarray}
with dimensions in angular frequency, $\omega$, and azimuthal mode number, $m$, and with dimensional lengths, $N_t$ and $N_\theta$, respectively.  Note that the coordinates $\omega=2 \pi f$ and $m$ range between their negative and positive Nyquist frequencies (-87.5~kHz $ \leq f <$ 87.5~kHz and $-N_\theta/2 \leq m < N_\theta/2$). Due to symmetry, the negative frequencies can be truncated.   

\begin{figure}
	\includegraphics{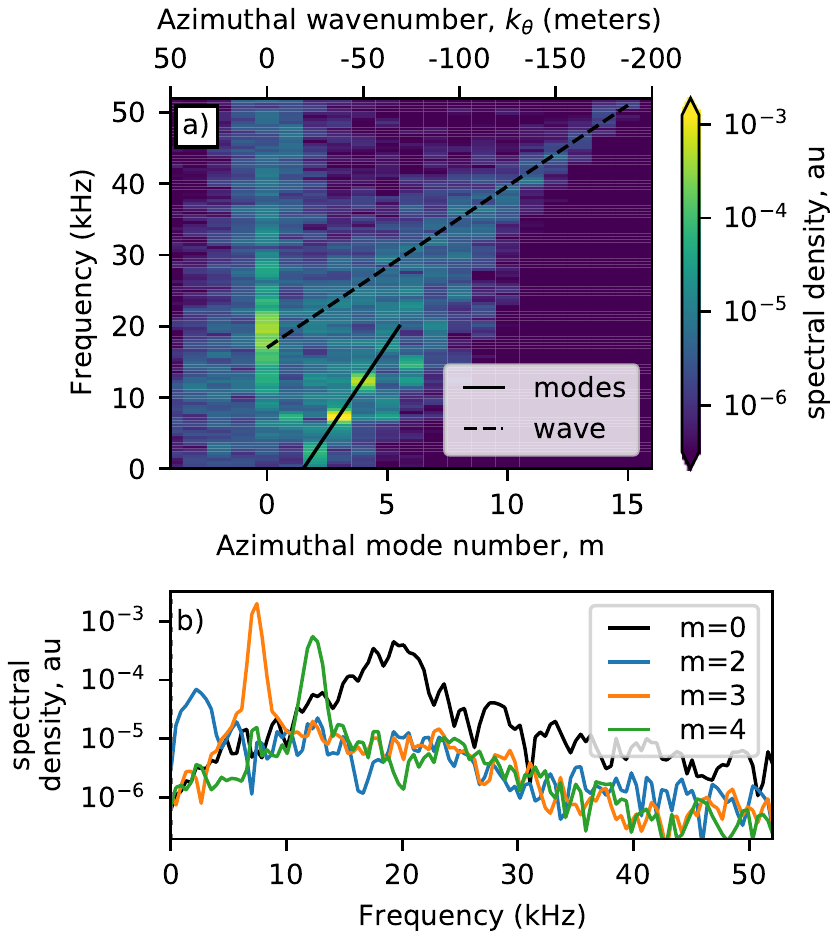}
	\caption{\label{fig:dispersion} a) The azimuthal dispersion plot within the channel.  A continuous wave and series of azimuthal mode numbers are identified and travel counter-clockwise (i.e. are negative).  b) The power spectrum of select mode numbers helps to identify their characteristic frequency.  }
\end{figure}

The absolute value of this result, $\bigl |P(\omega,m) \bigr|$, is the azimuthal dispersion relationship within the channel and is shown in Figure~\ref{fig:dispersion}a.  It reveals a series of evenly-spaced discrete modes ($2 \leq m \leq 5$) and what appears to be a continuous wave.  Both the continuous wave and the discrete modes are propagating clockwise (negative $\theta$) and therefore have negative wavenumbers.  By convention, we present the azimuthal modes and wave numbers as positive.   

Figure~\ref{fig:dispersion}b shows several slices of the dispersion relationship at $m=0$, 2, 3, and 4 and more clearly identifies the peaks originally observed in Figure~\ref{fig:raw_data}b.  From this plot, the $m=0$ mode is the broad-spectrum peak at 20 kHz, the $m=3$ mode is the narrow peak at 7.5 kHz, and the $m=4$ mode is the peak at 12.5 kHz.  
In addition, a weak $m=2$ mode is observed at 2.5 kHz. 
Figure~\ref{fig:dispersion}b shows that all three of the indicated $m>0$ modes have a roughly uniform spacing of 5 kHz.  Figure~\ref{fig:dispersion}a also identifies an $m=6$ mode at roughly double the $m=3$ frequency which makes it a harmonic of the $m=3$ mode~\cite{yamada2010}.  To observe the 80 kHz peak in Figure~\ref{fig:raw_data}b, the above analysis could be applied to the region around the cathode instead of inside the channel.

\subsection{Mode evolution \label{sec:fft_time}}

\begin{figure*}
	\includegraphics{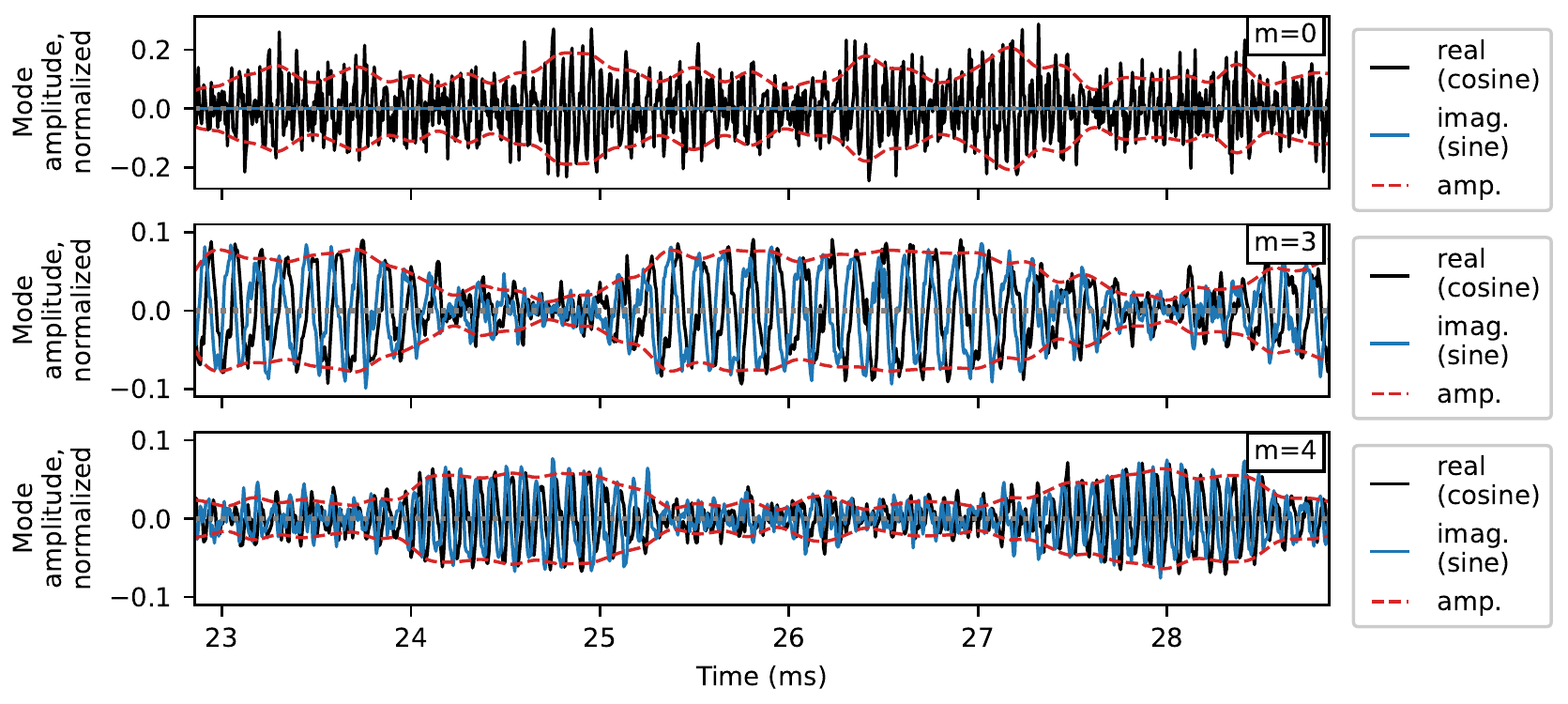}
	\caption{\label{fig:FFT_evolution} Time evolution of the $m=0$, 3 and 4 modes as captured by the 1D FFT in $\theta$.  Repeated mode hopping is observed between the $m=3$ and $m=4$ modes.  The signals are normalized by the average channel brightness.}
\end{figure*}

To capture the time evolution of each mode, we apply 1D FFT, $\mathcal{F}_{\theta}$, to each time step in the 2D dataset, $p(t,\theta)$, along the $\theta$ dimension.  The result is the 2D complex matrix, 

\begin{eqnarray}
	P(t,m)=\mathcal{F}_{\theta} \bigl \{p(t,\theta) \bigr \} 
	\label{eq:FFT_1D}
\end{eqnarray}
with dimensions in time, $t$, and azimuthal mode number, $m$.  Depending on the FFT algorithm, $P(t,m)$ then needs to be multiplied by a constant to return the correct amplitude, typically $2/N_\theta$. 
Figure~\ref{fig:FFT_evolution} plots the real (cosine) component, the imaginary (sine) component, and the amplitude of $P(t,m)$ at $m=0$, 3, and 4 and shows the time evolution of each.  
Figure~\ref{fig:FFT_evolution}a reveals the $m=0$ breathing mode to have a mostly consistent amplitude around 10\% to 20\% of the average channel brightness.  Figures~\ref{fig:FFT_evolution}b and~\ref{fig:FFT_evolution}c shows the $m=3$ and $m=4$ modes, respectively, with amplitudes between 3\% to 5\%.  These figures also clearly identify repeated mode hopping between these two modes.   In the $m=3$ and 4 figures, the real component leads the real component by roughly 90 degrees, indicating that the modes are rotating clockwise.   The $m=0$ mode is not rotating, and therefore its imaginary component is zero.  

\subsection{The modes' spatial structures}

\begin{figure}
	\includegraphics{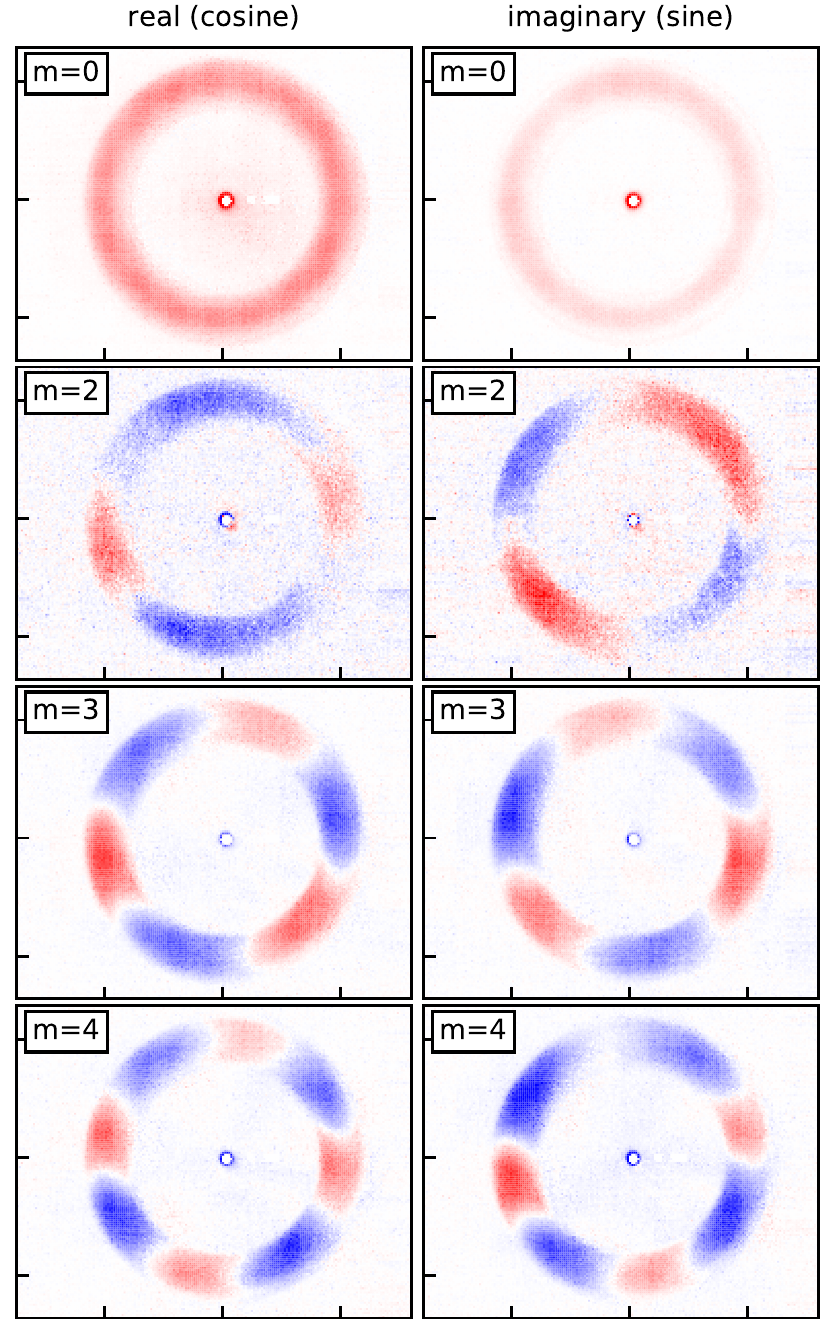}
	\caption{\label{fig:fft_spatial}
		Fourier analysis (Eq.~\ref{eq:FFT_2D}) reveals the spatial structure of several dominant modes in the channel.  The amplitude is spectral density (au) with each subfigure having a different scaling.   }
\end{figure}

To isolate the spatial structure associated with each mode, we first apply 1D Fourier analysis,

\begin{eqnarray}
	P(f,x,y)=\mathcal{F}_{t} \bigl \{p(t,x,y) \bigr \},
\end{eqnarray}
to each pixel in the 3D video, $p(t,x,y)$, with respect to time.  Next, we index the resulting matrix, $P(f,x,y)$, at the frequencies associated with each mode as identified in Figure~\ref{fig:dispersion}.  Figure~\ref{fig:fft_spatial} shows the resulting real (cosine) and imaginary (sine) component of each mode, which reveals the spatial extent and phase of each.  The number of peaks and troughs of each structure accurately corresponds to its mode number.  This approach is a computationally efficient alternative to bandpass-filtering each pixel as done previously~\cite{mcdonald2013}.  These plots are also nearly identical to plots in previous works~\cite{jorns2014, bairdthesis} with the difference being that the past works plot the phase, atan(imag./real), of each mode instead of the real and imaginary components.

\begin{figure}
	\includegraphics{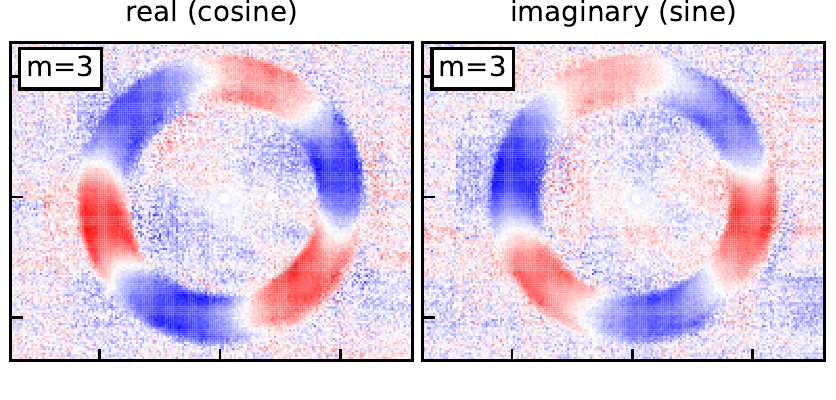}
	\caption{\label{fig:fft_spatial_normalized}
		Pre-normalizing each pixel in $p(t,x,y)$ by its standard deviation instead of by the average channel brightness provides better visualization of the mode structure.  Amplitude is spectral density (au).  These plots are directly comparable to the $m=3$ subplots in Figure~\ref{fig:fft_spatial} where the channel normalization is used.  }
\end{figure}

For better visualization, each pixel in $p(t,x,y)$ can optionally be normalized by its standard deviation before applying Eq.~\ref{eq:FFT_2D}.  Figure~\ref{fig:fft_spatial_normalized} shows an example of this for the $m=3$ mode.  This figure reveals improved spatial detail within the channel and also suggests mode structure outside the channel as well.

\subsection{Fourier methods conclusion}

In this section, we applied the Fourier mode decomposition and identification methods to our H6 high-speed video dataset.  First, the FFT of individual pixels was first able to identify the presence of coherent mode dynamics in the channel and cathode regions.  After radially and azimuthally binning the data within the channel, a 2D FFT provided an azimuthal dispersion plot and helped relate mode numbers to their characteristic frequency.  A 1D FFT was then used to isolate the temporal evolution of each mode and most notably identified mode hopping between the $m=3$ and $m=4$ modes in the channel.  Finally, a 1D FFT was applied to the original video, and plotting the frequencies associated with each provided the mode's spatial structure.  While these methods were only applied to the channel region in this work, this process can be applied to any radial region including around the cathode.  

\section{\label{sec:SVDmethod}SVD-based methods}

Matrix factorization-based methods, based on the SVD (Singular Value Decomposition) algorithm~\cite{Golub1970nm}, are an alternative approach to decomposing and identifying modes from high-speed video and provide several advantages over Fourier methods.  Below, we first detail the SVD algorithm, discuss the simplest SVD-based mode analysis method (the POD algorithm), and apply it to the present dataset.  

\subsection{\label{sec:SVD}The SVD algorithm}

The SVD algorithm is the foundation for POD and for more advanced algorithms.  This section discusses the SVD algorithm and its implementation.   

The most notable feature of SVD is that it does not presume spatial or temporal bases (e.g. Fourier sines and cosines) when decomposing coherent structures.  Instead, SVD provides a \textit{tailored}, orthonormal basis for a particular video dataset by minimizing the $L_2$ norm (error) between the original video and the video reconstructed from the new SVD bases~\cite{brunton2019}.  SVD orders these bases from the highest to lowest contribution to the reconstruction instead of by increasing wavenumber or frequency.  A downside to this is that the bases often need to be associated with their Fourier counterparts through post-processing (manually or algorithmically).  An advantage of SVD is that it does not require any preprocessing, most notably conversion to a particular coordinate system (e.g. converting the Cartesian video to polar).  
More details on the SVD algorithm can be found in the extensive literature on the subject~\cite{Golub1970nm,golub1996cf,woolfe2008fast,brunton2019}.

In order to apply the SVD algorithm to a video dataset, $p(t,x,y)$, we first convert it from 3D to 2D.  We do this by stacking the data associated with the two spatial dimensions into a single spatial dimension, $z=\text{stack}(x,y)$ to get the 2D $p(t,z)$ dataset with dimensions in time and space and with sizes $N_t$ and $N_z=N_x N_y$, respectively.  This process is the 3D equivalent of stacking columns in a 2D matrix to get a 1D array.  
The Python data structure \textit{xarray} has built-in functions (\textit{stack} and \textit{unstack}) that make stacking and unstacking very convenient.   
By convention, we also transpose $p(t,z)$ to $p(z,t)$ so that the spatial dimension is ordered before the temporal dimension.  For our high-speed video dataset,  $N_z>N_t$, so $p(z,t)$ is a “tall-skinny” matrix.  

When applied to $p(z,t)$, the SVD algorithm outputs three real matrices: $U(z,n)$, $\Sigma(n,n)$,  and $ V(t,n)^T $.  Here, we are using $n$ to represent the SVD basis numbers to distinguish them from the Fourier mode numbers, $m$.  First, $U(z,n)$ is a non-square, 2D matrix associated with the spatial bases (called the topos, i.e., ``shapes'') with dimensions of space, $z$, and mode number, $n$, and lengths $N_z$ and $N_n=N_t$, respectively.  $\Sigma(n,n)$ is a square 2D diagonal matrix associated with the basis energies (these diagonal values are referred to as the singular values) with dimensions $n$ by $n$ each with length $N_n$.  Finally, the transposed matrix, $V(t,n)^T$, is a square 2D matrix that contains the temporal evolution of the bases (called the chronos, i.e., ``times'') with dimensions of $n$ and $t$, each with length, $N_n=N_t$.  The exact dimension and format of these three matrices may vary depending on the particular SVD algorithm being used and shape of $p(z,t)$. 
For example, many SVD algorithms
return $\Sigma(n,n)$ as 1D array of its diagonal elements.

Multiplying these three matrices together, 
\begin{equation}
	\setlength{\arraycolsep}{1.5pt}
	\begin{split}
		&U(z,n)\Sigma(n,n)V(t,n)^T  \\
		&=\begin{bmatrix}
			| & | &  & | \\ 
			u_1 & u_2 & \ldots & u_N \\ 
			| & | &  & |
		\end{bmatrix} 
		\begin{bmatrix}
			\sigma_1 &  &   \\ 
			& \ddots &  \\ 
			&  & \sigma_N
		\end{bmatrix} 
		\begin{bmatrix}
			| & | &  & | \\ 
			v_1 & v_2 & \ldots & v_N \\ 
			| & | &  & |
		\end{bmatrix} ^T \\
		& = \sigma_1 u_1 v_1^T +  \sigma_2 u_2 v_2^T + \ldots + \sigma_N u_N v_N^T  \\
		& \approx p(z,t),
		\label{eq:svd_recon}
	\end{split}
\end{equation}

	\noindent reconstructs $p(z,t)$ with remarkable accuracy (owing to the orthonormality of the $u_n$ and $v_n$).  Related to this, another major advantage of SVD over Fourier is its ability to reconstruct data (like this high-speed video) with fewer bases.  In this equation, $u_n$, $\sigma_n$, and $v_n$, are the topo, singular value (energy), and chrono associated with the $n^{\mathrm{th}}$ basis, which are ordered from highest contribution (energy) to lowest. In Eq.~\ref{eq:svd_recon}, the multiplication of $u_n v_n^T$ is an outer product.  As a final note, we refer to $\sigma_n$ as the energy because our measurement is light intensity.  However, it is common in the literature to measure the velocity or magnetic field and therefore refer to $\sigma_n^2$ as the energy.

	\subsection{\label{sec:POD}Proper Orthogonal Decomposition (POD)}
	
	The POD method, also referred to as Biorthogonal Decomposition (BD), has been used extensively for plasma physics datasets~\cite{dudok1994} but has only been very recently introduced to Hall thruster and hollow cathode high-speed video
	~\cite{desangles2020,Becatti2021}.  
	The POD algorithm is nearly identical to the SVD algorithm with the main distinction being that the $U$, $\Sigma$, and $V^T$ matrices are often trimmed to retain a smaller set of bases (e.g. n<30).  
	
\begin{figure}
		\includegraphics{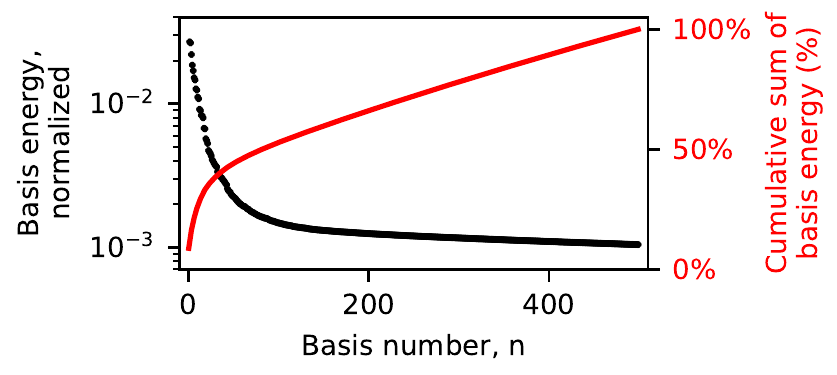}
		\caption{\label{fig:SVD_energy}
			The SVD algorithm orders its bases from highest to lowest energy (i.e. their total contribution to the reconstructed video), and therefore the lower numbered bases are more likely to contain the most prominent dynamics.   }
	\end{figure}

	After applying the POD algorithm (Section~\ref{sec:SVD}) to our dataset, we inspect each of the $U$, $\Sigma$, and $V^T$ matrices to identify and interpret the dynamics associated with each mode.  First, the energy for each basis, $\sigma_n$, is plotted in Figure~\ref{fig:SVD_energy}.  This highlights that the lower numbered bases capture the majority of the dynamics of the video and are therefore the focus of this section.  

\begin{figure}
		\includegraphics{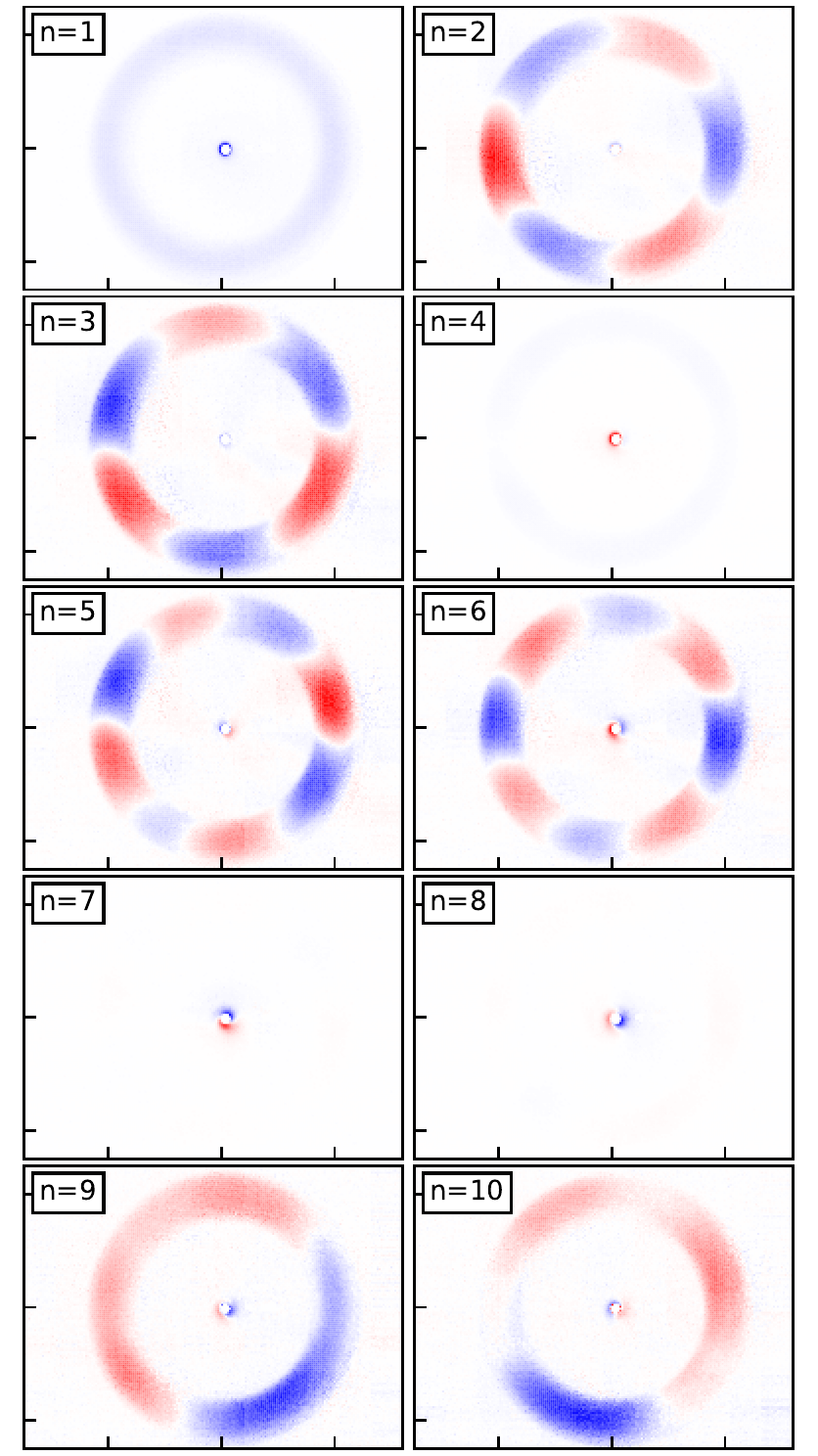}
		\caption{\label{fig:SVD_topos} The POD topos for the first 10 bases show $m=0$ or $m>0$ azimuthal structures in the channel or around the cathode.  Amplitude units are arbitrary.  By inspection, topos correspond to Fourier sine/cosine pairs as shown in Table~\ref{tab:mapping}}
	\end{figure}

	To visualize the spatial bases (topos), we unstack the $z$ dimension in $U(z,n)$ to get $U(x,y,n)$.  Figure~\ref{fig:SVD_topos} shows the first 10 topos, $u_n$, and each reveals an $m=0$ or $m>0$ azimuthal structure in the channel or around the cathode.  The bases associated with $m>0$ rotating modes each have a near-duplicate (but rotated) topo that represents an effective sine-cosine pairing.  For example, the $n=2$ and 3 bases are effectively the Fourier sine and cosine (real and imaginary) components of the $m=3$ rotating mode in the channel and therefore well matches the modes in Figure~\ref{fig:fft_spatial}.  Note that the POD bases associated with the $m=2$ Fourier mode is not shown in Figure~\ref{fig:SVD_topos} despite being present (as shown by Figure~\ref{fig:fft_spatial}).  This is because its energy contribution is lower than the first 10 POD bases.  
	
\begin{figure*}
		\includegraphics{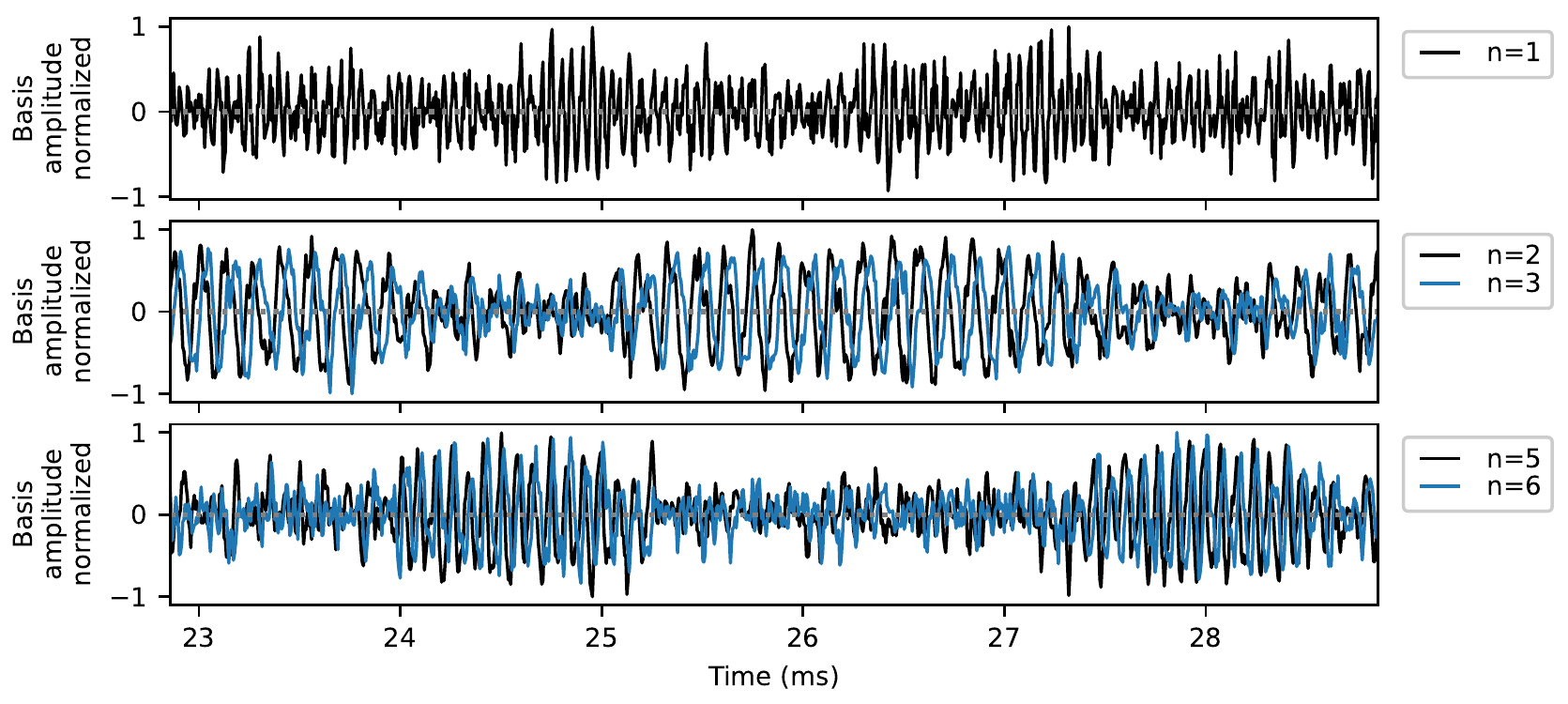}
		\caption{\label{fig:SVD_chronos} The POD chronos for the $n=1$, 2, 3, 5 and 6 bases.  These bases are roughly equivalent to the time evolutions of the $m=0$, 3, and 4 Fourier modes in Figure~\ref{fig:FFT_evolution}.  }
	\end{figure*}

	Each of these topos has an associated time evolution, i.e. chronos, associated with it.  Several select chronos, $\sigma_n v^T_n  $, are plotted in Figure~\ref{fig:SVD_chronos}.  Referring back to Figure~\ref{fig:SVD_topos}, we see that their corresponding topos are roughly equivalent to the $m=0$, 3, and 4 Fourier modes in the channel.  Therefore, it is not a surprise that their chronos in Figure~\ref{fig:SVD_chronos} are nearly identical to the Fourier-solved mode evolution shown in Figure~\ref{fig:FFT_evolution}.    
	
\begin{figure}
		\includegraphics{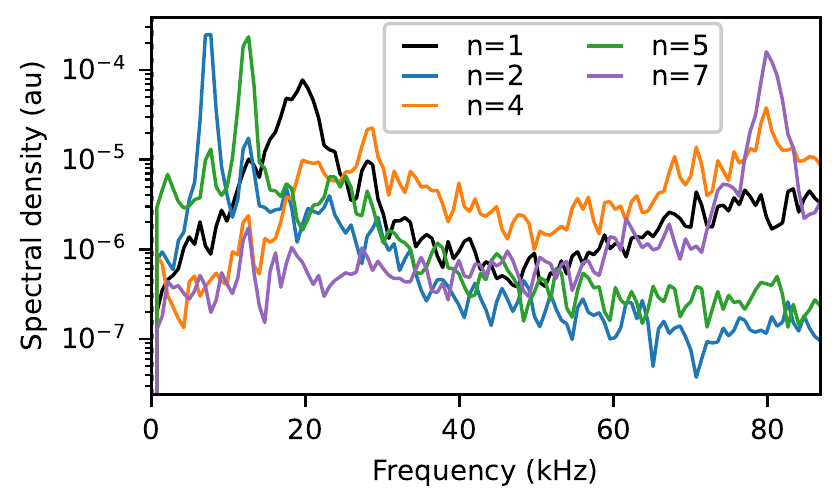}
		\caption{\label{fig:SVD_chronos_fft} The Welch-averaged FFT of select POD chronos helps in identifying the presence of a coherent mode structure and their characteristic frequencies.   This figure is directly comparable to the Fourier results in Figure~\ref{fig:dispersion}b. }
	\end{figure}

	Finally, the Welch-averaged FFT of select POD chronos are shown in Figure~\ref{fig:SVD_chronos_fft}.  Note that only a single basis from a sine-cosine pair is presented as their power spectrums of each are nearly identical.  Many of the peaks in Figure~\ref{fig:SVD_chronos_fft} are the same peaks as identified in the raw data (Figure~\ref{fig:raw_data}b) and the Fourier mode analysis (Figure~\ref{fig:dispersion}b).  A new structure is revealed by the $n=4$ basis, related to the Fourier $m=1$ cathode mode, which has two peaks around 38 and 80 kHz.  From this analysis alone, it is not clear if the $m=1$ cathode mode truly has two characteristic frequencies, if this basis is the combination of two Fourier modes, or if there is another explanation.  
	
	Comparing the results of the POD bases (Figures~\ref{fig:SVD_topos}-\ref{fig:SVD_chronos_fft}) to the Fourier modes (Figures~\ref{fig:dispersion}-\ref{fig:fft_spatial}) reveals a near one-to-one mapping as is indicated in Table~\ref{tab:mapping}.  This is not too surprising as there are already established scenarios in which POD modes reduce to Fourier modes~\cite{holmes2012turbulence}.

	\Table{\label{tab:mapping}The POD bases and Fourier modes show a near one-to-one mapping.} 
	\br
	\textbf{POD basis, n}&\textbf{Fourier mode, m}&\textbf{Location}\\
	\mr
	1		& 	0					& 	Channel\\
	2, 3	& 	3, sine and cosine	& 	Channel\\
	4		& 	0					& 	Cathode\\
	4, 6	& 	4, sine and cosine	& 	Channel\\
	7, 8 	& 	1, sine and cosine	& 	Cathode\\
	\br
	\endTable

\subsection{SVD methods conclusion}

The primary advantage of the SVD-based methods is that they do not presume a universal basis, Fourier or otherwise, and instead create tailored bases which optimally capture particular features of the dynamics.  In addition, these methods require less preconditioning of the video, e.g. radial binning and converting the pixel coordinates to polar, and therefore they can provide detailed images of the spatial structures of each basis or mode (Figure~\ref{fig:SVD_topos}).   The most basic SVD method, POD, is easier to implement than the Fourier methods, does not require a linear assumption, and produces very similar results.  
More advanced SVD methods, such as DMD, have potential for Hall thruster modal analysis but require more development.

\section{\label{sec:conclusion}Conclusions}

This work provided a survey of existing Fourier and new SVD-based mode decomposition techniques for high-speed Hall thruster video data.  To highlight the implementation of these methods, they were each applied to the same H6 dataset which contained simultaneous modes associated with the channel and cathode.  Both Fourier and POD methods were able to characterize the spatial and temporal evolution of each mode, including mode hopping between the $m=3$ and $m=4$ channel modes.  Both methods also had their various strengths and weaknesses.

Fourier methods are well suited for characterizing linear wave dynamics and therefore excel at identifying both discrete mode and continuous wave dynamics in the various regions of the Hall thruster plasma.  A notable feature of the Fourier methods are their ability to naturally provide dispersion relationships.  However, Fourier methods require preprocessing, e.g. converting the video to polar coordinates,  which adds computational complexity.  

The main advantage of the SVD methods (notably POD) is that they produce a tailored set of orthogonal bases for each dataset that are ordered by their relative amplitudes (high to low), and these bases do not presume any form (e.g. sines and cosines in Fourier).  The POD method is relatively simple to implement as preprocessing, most notably coordinate conversion, is optional.  In addition, the POD method can naturally locate mode structures at different spatial locations (e.g. cathode and anode) unlike Fourier which requires the dataset to be preprocessed with a particular spatial location in mind.  

Together, Fourier and POD methods provide for a more complete toolkit for studying global Hall thruster dynamics as captured by high-speed imaging.  Moving forward, we recommend first applying POD to a dataset in order to identify its major dynamics and the spatial locations of each.  With this information, Fourier methods can then be intelligently targeted to extract dispersion relationships or other select modal information.  The code and data used in this work are made are available online~\cite{zenodo_2021} to assist future researchers.

\section*{\label{sec:acknowledgments}Acknowledgments}

This analysis was performed while JWB held an NRC Research Associateship award at the Naval Research Laboratory (NRL), with JWB and MM supported by the NRL Base Program. The imaging data was collected at the University of Michigan by MM under advisor Alec Gallimore in the Plasmadynamics and Electric Propulsion Laboratory, where the use of the Photron SA5 FASTCAM was made possible via grant FA9550-09-1-0695 from the Air Force of Scientific Research (AFOSR). AAK acknowledges funding from the Army Research Office ({ARO W}911{NF}-19-1-0045) and the Air Force Office of Scientific Research (AFOSR {FA}9550-18-1-0200).


\section*{References}
\nocite{*}
\bibliographystyle{iopart-num}
\bibliography{bibliography}%

\providecommand{\noopsort}[1]{}\providecommand{\singleletter}[1]{#1}%
\providecommand{\newblock}{}
\begin{thebibliography}{10}
\expandafter\ifx\csname url\endcsname\relax
  \def\url#1{{\tt #1}}\fi
\expandafter\ifx\csname urlprefix\endcsname\relax\def\urlprefix{URL }\fi
\providecommand{\eprint}[2][]{\url{#2}}

\bibitem{goebel2008}
Goebel D~M and Katz I 2008 {\em Fundamentals of Electric Propulsion: Ion and
  Hall Thrusters\/} ({\em {JPL} Space Science And Technology Series\/} no~1)
  (NASA JPL)
  \urlprefix\url{https://www.wiley.com/en-us/Fundamentals+of+Electric+Propulsion%3A+Ion+and+Hall+Thrusters-p-9780470436264}

\bibitem{boeuf2017}
Boeuf J~P 2017 {\em Journal of Applied Physics\/} {\bf 121} 011101 ISSN
  0021-8979
  \urlprefix\url{https://aip-scitation-org.ezproxy.cul.columbia.edu/doi/10.1063/1.4972269}

\bibitem{florenz2012}
Florenz R, Liu T, Gallimore A, Kamhawi H, Brown D, Hofer R and Polk J Electric
  propulsion of a different class: The challenges of testing for {MegaWatt}
  missions {\em 48th {AIAA}/{ASME}/{SAE}/{ASEE} Joint Propulsion Conference \&
  Exhibit\/} Joint Propulsion Conferences (AIAA)
  \urlprefix\url{https://arc.aiaa.org/doi/10.2514/6.2012-3942}

\bibitem{hall2017}
Hall S, Jorns B, Gallimore A, Kamhawi H, Haag T, Mackey J, Gilland J, Peterson
  P and Baird M High-power performance of a 100-{kW} class nested hall thruster
  {\em 35th International Electric Propulsion Conference\/}
  \urlprefix\url{https://pepl.engin.umich.edu/pdf/IEPC-2017-228.pdf}

\bibitem{choueiri2001}
Choueiri E~Y 2001 {\em Physics of Plasmas\/} {\bf 8} 1411--1426 ISSN 1070-664X
  \urlprefix\url{https://aip.scitation.org/doi/abs/10.1063/1.1354644}

\bibitem{parker2010}
Parker J~B, Raitses Y and Fisch N~J 2010 {\em Applied Physics Letters\/} {\bf
  97} 091501 ISSN 0003-6951 \urlprefix\url{https://doi.org/10.1063/1.3486164}

\bibitem{mcdonald2011segAnode}
McDonald M~S and Gallimore A~D 2011 Measurement of {Cross}-{Field} {Electron}
  {Current} in a {Hall} {Thruster} {Due} to {Rotating} {Spoke} {Instabilities}
  {\em 47th {AIAA}/{ASME}/{SAE}/{ASEE} {Joint} {Propulsion} {Conference} \&
  {Exhibit}\/} (AIAA 2011-5810)
  \urlprefix\url{https://arc.aiaa.org/doi/abs/10.2514/6.2011-5810}

\bibitem{jorns2020}
Jorns B~A, Cusson S~E, Brown Z and Dale E 2020 {\em Physics of Plasmas\/} {\bf
  27} 022311 ISSN 1070-664X
  \urlprefix\url{https://aip.scitation.org/doi/full/10.1063/1.5130680}

\bibitem{sekerak2016}
Sekerak M~J, Gallimore A~D, Brown D~L, Hofer R~R and Polk J~E  {\bf 32}
  903--917 \urlprefix\url{https://doi.org/10.2514/1.B35709}

\bibitem{dale2019}
Dale E Two-zone hall thruster breathing mode mechanism, part {I}: Theory {\em
  36th International Electric Propulsion Conference\/}
  \urlprefix\url{https://pepl.engin.umich.edu/pdf/IEPC-2019-354.pdf}

\bibitem{goebel2007}
Goebel D~M, Jameson K~K, Katz I and Mikellides I~G  {\bf 14} 103508 ISSN
  1070-664X \urlprefix\url{https://aip.scitation.org/doi/10.1063/1.2784460}

\bibitem{georgin2019}
Georgin M~P, Jorns B~A and Gallimore A~D  {\bf 26} 082308 ISSN 1070-664X
  \urlprefix\url{https://aip.scitation.org/doi/10.1063/1.5111552}

\bibitem{jorns2014}
Jorns B~A and Hofer R~R 2014 {\em Physics of Plasmas\/} {\bf 21} 053512 ISSN
  1070-664X \urlprefix\url{https://doi.org/10.1063/1.4879819}

\bibitem{janes1966}
Janes G~S and Lowder R~S 1966 {\em The Physics of Fluids\/} {\bf 9} 1115--1123
  ISSN 0031-9171
  \urlprefix\url{https://aip.scitation.org/doi/10.1063/1.1761810}

\bibitem{mcdIEEE-TPSrotspoke}
McDonald M~S and Gallimore A~D 2011 {\em IEEE Transactions on Plasma Science\/}
  {\bf 39} 2952--2953 ISSN 0093-3813
  \urlprefix\url{http://umich.edu/~peplweb/pdf/IEEE-TransPlasmaSci-2161343-2011.pdf}

\bibitem{bairdthesis}
Baird M 2020 {\em Investigating {Newly} {Discovered} {Oscillation} {Modes} in
  {Magnetically} {Shielded} {Hall} {Effect} {Thrusters} {Utilizing} {High}
  {Speed} {Diagnostics}\/} Ph.D. thesis Western Michigan University Kalamazoo,
  MI \urlprefix\url{https://scholarworks.wmich.edu/dissertations/3656}

\bibitem{mcdonald2013}
McDonald M~S and Gallimore A~D 2013 Comparison of {Breathing} and {Spoke}
  {Mode} {Strength} in the {H6} {Hall} {Thruster} {Using} {High} {Speed}
  {Imaging} {\em 33rd {International} {Electric} {Propulsion} {Conference}\/}
  (Washington, DC) \urlprefix\url{http://electricrocket.org/IEPC/28xa7w8t.pdf}

\bibitem{darnon1997}
Darnon F, Lyszyk M, Bouchoule A, Darnon F, Lyszyk M and Bouchoule A 1997
  Optical investigation on plasma investigations of {SPT} thrusters {\em 33rd
  {Joint} {Propulsion} {Conference} and {Exhibit}\/} (American Institute of
  Aeronautics and Astronautics)
  \urlprefix\url{https://arc.aiaa.org/doi/abs/10.2514/6.1997-3051}

\bibitem{hara2014}
Hara K, Sekerak M~J, Boyd I~D and Gallimore A~D 2014 {\em Journal of Applied
  Physics\/} {\bf 115} 203304 ISSN 0021-8979
  \urlprefix\url{https://aip-scitation-org.ezproxy.cul.columbia.edu/doi/10.1063/1.4879896}

\bibitem{mcdonald2011spokeUbiquity}
McDonald M and Gallimore A 2011 Parametric investigation of the rotating spoke
  instability in hall thrusters {\em 32nd {International} {Electric}
  {Propulsion} {Conference}\/} (Wiesbaden, Germany: IEPC 2011-242)
  \urlprefix\url{http://www.umich.edu/~peplweb/pdf/IEPC-2011-242.pdf}

\bibitem{romadanov2019}
Romadanov I, Raitses Y and Smolyakov A 2019 {\em Plasma Physics Reports\/} {\bf
  45} 134--146 ISSN 1562-6938
  \urlprefix\url{https://doi.org/10.1134/S1063780X19020156}

\bibitem{brunton2019}
Brunton S~L and Kutz J~N 2019 {\em Data-{Driven} {Science} and {Engineering}:
  {Machine} {Learning}, {Dynamical} {Systems}, and {Control}\/} 1st ed
  (Cambridge: Cambridge University Press) ISBN 978-1-108-42209-3
  \urlprefix\url{https://doi.org/10.1017/9781108380690}

\bibitem{Benner2015siamreview}
Benner P, Gugercin S and Willcox K 2015 {\em SIAM review\/} {\bf 57} 483--531
  \urlprefix\url{https://epubs.siam.org/doi/10.1137/130932715}

\bibitem{Taira2017aiaa}
Taira K, Brunton S~L, Dawson S, Rowley C~W, Colonius T, McKeon B~J, Schmidt
  O~T, Gordeyev S, Theofilis V and Ukeiley L~S 2017 {\em AIAA Journal\/} {\bf
  55} 4013--4041 \urlprefix\url{https://arc.aiaa.org/doi/10.2514/1.J056060}

\bibitem{noack2003hierarchy}
Noack B~R, Afanasiev K, Morzy{\'n}ski M, Tadmor G and Thiele F 2003 {\em
  Journal of Fluid Mechanics\/} {\bf 497} 335--363
  \urlprefix\url{https://doi.org/10.1017/S0022112003006694}

\bibitem{oudheusden2005}
Oudheusden B, Scarano F, Van~Hinsberg N and Watt D 2005 {\em Experiments in
  Fluids\/} {\bf 39} 86--98
  \urlprefix\url{https://link.springer.com/article/10.1007/s00348-005-0985-5}

\bibitem{dudok1994}
Dudok~de Wit T, Pecquet A, Vallet J and Lima R 1994 {\em Physics of Plasmas\/}
  {\bf 1} 3288--3300 ISSN 1070-664X
  \urlprefix\url{https://aip.scitation.org/doi/10.1063/1.870481}

\bibitem{levesque2013}
Levesque J~P, Rath N, Shiraki D, Angelini S, Bialek J, Byrne P~J, DeBono B~A,
  Hughes P~E, Mauel M~E, Navratil G~A, Peng Q, Rhodes D~J and Stoafer C~C 2013
  {\em Nuclear Fusion\/} {\bf 53} 073037 ISSN 0029-5515
  \urlprefix\url{https://iopscience.iop.org/article/10.1088/0029-5515/53/7/073037}

\bibitem{vanMilligen14}
van Milligen B~P, S{\'{a}}nchez E, Alonso A, Pedrosa M~A, Hidalgo C,
  de~Aguilera A~M and Fraguas A~L 2014 {\em Plasma Physics and Controlled
  Fusion\/} {\bf 57} 025005
  \urlprefix\url{https://iopscience.iop.org/article/10.1088/0741-3335/57/2/025005/meta}

\bibitem{hansen2015numerical}
Hansen C, Victor B, Morgan K, Jarboe T, Hossack A, Marklin G, Nelson B and
  Sutherland D 2015 {\em Physics of Plasmas\/} {\bf 22} 056105
  \urlprefix\url{https://doi.org/10.1063/1.4919277}

\bibitem{kaptanoglu2020physics}
Kaptanoglu A~A, Morgan K~D, Hansen C~J and Brunton S~L 2021 {\em Phys. Rev.
  E\/} {\bf 104}(1) 015206
  \urlprefix\url{https://link.aps.org/doi/10.1103/PhysRevE.104.015206}

\bibitem{desangles2020}
Désangles V, Shcherbanev S, Charoy T, Clément N, Deltel C, Richard P, Vincent
  S, Chabert P and Bourdon A 2020 {\em Atmosphere\/} {\bf 11} 518
  \urlprefix\url{https://www.mdpi.com/2073-4433/11/5/518}

\bibitem{Becatti2021}
Becatti G, Goebel D~M and Zuin M  {\bf 129} 033304 ISSN 0021-8979, 1089-7550
  \urlprefix\url{http://aip.scitation.org/doi/10.1063/5.0028566}

\bibitem{schmid2010dynamic}
Schmid P~J 2010 {\em Journal of Fluid Mechanics\/} {\bf 656} 5--28
  \urlprefix\url{https://doi.org/10.1017/S0022112010001217}

\bibitem{Tu2014jcd}
Tu J~H, Rowley C~W, Luchtenburg D~M, Brunton S~L and Kutz J~N 2014 {\em Journal
  of Computational Dynamics\/} {\bf 1} 391--421
  \urlprefix\url{https://www.aimsciences.org/article/doi/10.3934/jcd.2014.1.391}

\bibitem{taylor2018dynamic}
Taylor R, Kutz J~N, Morgan K and Nelson B~A 2018 {\em Review of Scientific
  Instruments\/} {\bf 89} 053501
  \urlprefix\url{https://doi.org/10.1063/1.5027419}

\bibitem{Sasaki2019}
Sasaki M, Kawachi Y, Dendy R~O, Arakawa H, Kasuya N, Kin F, Yamasaki K and
  Inagaki S 2019 {\em Plasma Physics and Controlled Fusion\/} {\bf 61} 112001
  \urlprefix\url{https://iopscience.iop.org/article/10.1088/1361-6587/ab471b}

\bibitem{kaptanoglu2020}
Kaptanoglu A~A, Morgan K~D, Hansen C~J and Brunton S~L 2020 {\em Physics of
  Plasmas\/} {\bf 27} 032108
  \urlprefix\url{https://aip.scitation.org/doi/10.1063/1.5138932}

\bibitem{zenodo_2021}
Brooks J 2021 Code and data for a comparison of fourier and pod mode
  decomposition methods for high-speed hall thruster video
  \urlprefix\url{http://dx.doi.org/10.5281/zenodo.5150716}

\bibitem{brown2009}
Brown D~L 2009 {\em Investigation of Low Discharge Voltage Hall Thruster
  Characteristics and Evaluation of Loss Mechanisms.\/} Ph.D. thesis University
  of Michigan
  \urlprefix\url{https://deepblue.lib.umich.edu/handle/2027.42/63660}

\bibitem{kasa1976}
Kåsa I 1976 {\em IEEE Transactions on Instrumentation and Measurement\/} {\bf
  IM-25} 8--14 ISSN 1557-9662
  \urlprefix\url{https://ieeexplore.ieee.org/document/6312298}

\bibitem{taubin1991}
Taubin G 1991 {\em IEEE Transactions on Pattern Analysis and Machine
  Intelligence\/} {\bf 13} ISSN 1939-3539
  \urlprefix\url{https://ieeexplore.ieee.org/document/103273}

\bibitem{vora1997}
Vora P~L, Farrell J~E, Tietz J~D and Brainard D~H 1997 Linear models for
  digital cameras {\em Proceedings, IS\&T's 50th Annual Conference\/} pp
  377--382
  \urlprefix\url{http://citeseerx.ist.psu.edu/viewdoc/summary?doi=10.1.1.53.7490}

\bibitem{mcdonald2011}
McDonald M, Bellant C, St~Pierre B and Gallimore A 2011 Measurement of
  cross-field electron current in a hall thruster due to rotating spoke
  instabilities {\em 47th {AIAA} {Joint} {Prop}. {Conf}.\/}
  \urlprefix\url{https://arc.aiaa.org/doi/abs/10.2514/6.2011-5810}

\bibitem{welch1967}
Welch P 1967 {\em IEEE Transactions on Audio and Electroacoustics\/} {\bf 15}
  ISSN 1558-2582 \urlprefix\url{https://doi.org/10.1109/TAU.1967.1161901}

\bibitem{yamada2010}
Yamada T, Itoh S~I, Inagaki S, Nagashima Y, Shinohara S, Kasuya N, Terasaka K,
  Kamataki K, Arakawa H, Yagi M, Fujisawa A and Itoh K 2010 {\em Physics of
  Plasmas\/} {\bf 17} ISSN 1070-664X
  \urlprefix\url{https://aip.scitation.org/doi/10.1063/1.3429674}

\bibitem{Golub1970nm}
Golub G~H and Reinsch C 1970 {\em Numerical Mathematics\/} {\bf 14} 403--420
  \urlprefix\url{https://link.springer.com/article/10.1007/BF02163027}

\bibitem{golub1996cf}
Golub G~H and Loan C~F~V 1996 {\em Matrix {{Computations}}\/} 3rd ed
  ({Baltimore}: {Johns Hopkins University Press}) ISBN 978-0-8018-5414-9
  \urlprefix\url{https://jhupbooks.press.jhu.edu/title/matrix-computations}

\bibitem{woolfe2008fast}
Woolfe F, Liberty E, Rokhlin V and Tygert M 2008 {\em Applied and Computational
  Harmonic Analysis\/} {\bf 25} 335--366
  \urlprefix\url{https://doi.org/10.1016/j.acha.2007.12.002}

\bibitem{holmes2012turbulence}
Holmes P, Lumley J~L, Berkooz G and Rowley C~W 2012 {\em Turbulence, coherent
  structures, dynamical systems and symmetry\/} (Cambridge university press)
  \urlprefix\url{https://doi.org/10.1017/CBO9780511919701}

\end{thebibliography}

\end{document}